\documentclass[english]{article}
\usepackage[T1]{fontenc}
\usepackage[latin9]{inputenc}
\usepackage{geometry}
\usepackage{bm}
\usepackage{amsmath}
\usepackage{amssymb}
\usepackage{graphicx}
\usepackage{esint}
\usepackage{comment}
\usepackage{stmaryrd}
\usepackage{url}

\begin{document}

\title{Curvilinear coordinate Generalized Source Method for gratings with sharp edges}

\author{
Alexey A. Shcherbakov,$^{1,2,\ast}$\\
$^{1}$The International Research Center Nanophotonics and Metamaterials,\\ ITMO University, 49 Kronverksky Pr., St. Petersburg, 197101, Russia\\
$^{2}$Laboratory of Nanooptics and Plasmonics,\\ Moscow Institute of Physics and Technology, Institutsky lane 9, Dolgopudny, 141700, Russia \\
$^{\ast}$alexey.shcherbakov@metalab.ifmo.ru
}

\maketitle

\begin{abstract}
High-efficient direct numerical methods are currently in demand for
optimization procedures in the fields of both conventional diffractive
and metasurface optics. With a view of extending the scope of application
of the previously proposed Generalized Source Method in the curvilinear
coordinates, which has theoretical $O\left(N\log N\right)$ asymptotic numerical complexity, a new method formulation is developed for gratings with sharp edges. It is shown that corrugation corners can be treated as effective medium interfaces within the rationale of the method. Moreover, the given formulation is demonstrated to allow for application of the same derivation as one used in classical electrodynamics to derive the interface conditions. This yields continuous combinations of the fields and metric tensor components, which can be directly Fourier factorized. Together with an efficient algorithm the new formulation is demonstrated to substantially increase the computation accuracy for given computer resources.
\end{abstract}

\section{Introduction}
Electromagnetic planar grating diffraction problem for resonant domain
structures composed of dispersive materials can be solved by various
numerical methods depending on particular structure properties \cite{Petit1980,Popov2012}.
While the direct diffraction problem often is not an issue in practice,
related inverse problems impose strict requirements on the mentioned
methods, as a huge number of direct problem simulation runs is generally
required within inversion procedures \cite{Rodriguez2018}. This makes
the research directed towards an increase of simulation efficiency
in terms of computing resource requirements be in demand in the field
of engineering of high-efficient optical structures, like resonant
gratings \cite{Vuckovic2018}. In particular, efficient optimization
procedures are of great interest within the modern trend of metasurface
optimization, which is aimed at pushing forward the field so as to
substitute conventional diffractive optical components with high-index
metastructures.

Fourier space methods are good candidates for the purpose of large
scale optimization of periodic dielectric structures being versatile
in terms of structure geometries and offering a reliable convergence
error analysis \cite{Lalanne1996,Granet1996}. There were proposed
several fast numerical schemes exhibiting an $O\left(N\log N\right)$
numerical complexity and $O\left(N\right)$ computational memory resort
\cite{Shcherbakov2010,Beurden2011,Shcherbakov2012,Skobelev2013,Brenner2018},
and providing a large room for parallelization on vector processors. This asymptotic behaviour is due to a specific structure of the scattering operator, which factorizes into block-diagonal and block-Toeplitz components, and due to the possibility to multiply Towplitz matrices by vectors by means of the Fast Fourier Transform (FFT).
Owing to the remarkable features, the named methods possess a great potential
for applications in optimization procedures, though, some additional
specific improvements are required to ensure convergence for complex
diffractive optical elements and metasurfaces. This includes the scattering vector algorithm \cite{Iff2017}, which implements
the divide-and-conquer strategy for thick structures, and application
of curvilinear coordinate transformations, which enhance the numerical
behaviour of the Fourier methods. These transformations are of two
types. The first type implies in-plane stretching and contracting
a grating, and was developed within the context of the Fourier Modal
Method (FMM) \cite{Granet1999,Granet2001,Weiss2009} and the Modal
Method (MM) \cite{Edee2013,Edee2016} to simplify the correct treatment
of the boundary conditions \cite{Li1996,Popov2001,Feldbacq2009} and
improve numerical solution conditioning. A similar procedure can be
incorporated into the fast methods, though this possibility is not
considered in this work. The second type of transformations, which
inevitably affects the coordinate orthogonal to the grating plane,
includes transformations converting a complex corrugation interface
to a plane. An idea of such transformations gave rise to the Chandezon
method \cite{Chandezon1980,Granet1998,Li2015}, which is a Fourier modal method in the curvilinear coordinate space \cite{Felix2014}. In \cite{Shcherbakov2013,Shcherbakov2017} this idea was expanded to fit the rationale of computationally efficient
methods by introducing the generalized metric sources. The corresponding
method will be referred to as the Generalized Source Method in Curvilinear
Coordinates (GSMCC), and its implementation paved a way to the efficient
Fourier space simulation of metallic structures. However, the formulation
given in \cite{Shcherbakov2013,Shcherbakov2017} is based on an assumption
of the smoothness of functions describing periodic corrugations, thus,
having a limited range of validity. In this work a generalization
of the latter methods is developed in case of 1D gratings, which profile
functions are allowed to have corners at a limited number of points, so that the profiles can be flattened with appropriate coordinate transformations. Note also, that the method described here represents a rigorous explicit numerical procedure of solving Maxwell's equations with a controlled accuracy, and its validity is not constrained by convergence limitations being inherent to explicit and perturbative techniques like \cite{Reitich1993}, or Rayleigh series based methods \cite{Tishchenko2009}.

The article structure is the following. The narrative starts with
the grating diffraction problem statement. Then the GSMCC rationale
is outlined in form of a sequence of steps leading to a required solution
form. Details on the corresponding derivations can be found in the
previous papers \cite{Shcherbakov2013,Shcherbakov2017,Shcherbakov2018}.
A consequent analysis of effective metric sources and the Maxwell's
equations in curvilinear metric yields the derivation of correct handling
of truncated Fourier vectors. Next, an efficient numerical method
based on these rules is described. Numerical examples and the discussion
of the results enclose the paper.

\section{Grating diffraction}
Let us consider a simple case of a periodic grating corrugation separating
two different media, which is supposed to be described by a piece-wise smooth periodic function with bounded derivative $x_{3}=f\left(x_{1}\right)$, with
period $\Lambda$, such that $f\left(x_{1}+n\Lambda\right)=f\left(x_{1}\right)$,
$n\in\mathbb{N}$, in the Cartesian coordinates $x_{\alpha}$, $\alpha=1,2,3$.
Without loss of generality one can suppose that $\min f\left(x_{3}\right)=-a$,
$\max f\left(x_{3}\right)=a$. Dielectric permittivities of the media
below and above the corrugation interface are denoted as $\varepsilon_{1}$
and $\varepsilon_{2}$ respectively. Fig. 1(a) illustrates the setup.
Linear electromagnetic optical diffraction is governed by the time-harmonic
Maxwell's equations with implicit time dependence factor $\exp\left(-i\omega t\right)$
\begin{equation}
\begin{array}{c}
	\xi_{\alpha\beta\gamma}\partial_{\beta}E_{\gamma}\left(\bm{r}\right) = i\omega\mu_{0}H_{\alpha}\left(\bm{r}\right)-M_{\alpha}\left(\bm{r}\right),\\
	\xi_{\alpha\beta\gamma}\partial_{\beta}H_{\gamma}\left(\bm{r}\right) = -i \omega\varepsilon\left(\bm{r}\right)E_{\alpha}\left(\bm{r}\right)+J_{\alpha}\left(\bm{r}\right)
\end{array}\label{eq:Maxw_cartesian}
\end{equation}
subject to the continuity of the tangential fields at the corrugation
interface, and the radiation condition at $x_{3}\rightarrow\pm\infty$
\cite{Popov2012}. Here $J_{\alpha}$, $M_{\alpha}$ are the electric and magnetic source terms respectively. The Greek indices vary in range $1,2,3$, and $\bm{r}=\left(x_{1},x_{2},x_{3}\right)^{T}$. The Levi-Civita symbol $\xi_{\alpha\beta\gamma}=1$ for even permutations of $(1,2,3)$, and $\xi_{\alpha\beta\gamma}=-1$ for odd permutations. Summation over repeated coordinate indices is supposed here and further. Permittivity function is $\varepsilon\left(\bm{r}\right)=\varepsilon_{1}$ when $x_{3}\leq f\left(x_{1}\right)$, and $\varepsilon\left(\bm{r}\right)=\varepsilon_{2}$ when $x_{3}>f\left(x_{1}\right)$. A generalization may include a substrate, such that $\varepsilon\left(\bm{r}\right)=\varepsilon_{3}$ for $x_{3}<b$.

To set up a general solution of Eqs. (\ref{eq:Maxw_cartesian}), first,
the dielectric permittivity and the magnetic permeability inside some
region $D_{g}=\left\{ \left|x_{3}\right|\leq b:b\geq a\right\} $
incorporating the grating are assumed to be constant everywhere $\varepsilon\left(\bm{r}\right)=\varepsilon_{b}$,
$\mu=\mu_{b}$, with the subscript standing for ``basis''. This
assumption replaces the grating with a homogeneous plane layer. The
electromagnetic field for this basis layer can be found for any sources
by means of the volume integral equations with the electric and magnetic Green's
tensors $G_{\alpha\beta}^{E,M}\left(\bm{r},\bm{r}'\right)$. Attributing
the ``difference'' between the grating and the layer to effective
(generalized) sources, one attains a self-consistent linear equation
system for the searched solution of the diffraction problem. For the
planar grating problem the Green's tensors are subject to the plane
wave decomposition (see, e.g., \cite{Tsang2000,Shcherbakov2018}).
In order to get rid of a singular term proportional to $\delta\left(x_{3}-x'_{3}\right)$ present in the explicit form of $G_{\alpha\beta}^{E}\left(\bm{r},\bm{r}'\right)$ it is convenient to introduce the modified fields as
\begin{equation}
\begin{array}{l}
	\tilde{E}_{3}=E_{3}-J_{3}/i\omega\varepsilon_{b} \\
	\tilde{H}_{3}=H_{3}-M_{3}/i\omega\mu_{b}
\end{array}
\label{eq:mod}
\end{equation}
with the rest components being untouched: $\tilde{E}_{1,2}=E_{1,2}$, $\tilde{H}_{1,2}=H_{1,2}$.
The periodicity allows one to fix a zero harmonic wave vector $\bm{k}^{inc}=\left(k_{1}^{inc},0,k_{3}^{inc}\right)$
and to represent the fields and sources in form of Bloch waves. This
makes possible a decomposition of the modified electromagnetic field
into a set of TE (superscript ``$e$'') and TM (superscript ``$h$'')
polarized plane waves propagating upwards and downwards relative to
the vertical coordinate $x_{3}$:
\begin{equation}
\tilde{E}_{\alpha}\left(\bm{r}\right)=\sum_{m=-\infty}^{\infty}\sum_{\sigma=\pm}\left[\tilde{a}_{m}^{e\sigma}\left(x_{3}\right)\hat{e}_{m\alpha}^{e\sigma}+\tilde{a}_{m}^{h\sigma}\left(x_{3}\right)\hat{e}_{m}^{h\sigma}\right]\exp\left(ik_{m1}x_{1}+i\sigma k_{m3}x_{3}\right),\label{eq:E_pw}
\end{equation}
\begin{equation}
\tilde{H}_{\alpha}\left(\bm{r}\right)=\frac{k_{b}}{\omega\mu_{0}}\sum_{m=-\infty}^{\infty}\sum_{\sigma=\pm}\left[\tilde{a}_{m}^{h\sigma}\left(x_{3}\right)\hat{e}_{m\alpha}^{e\sigma}-\tilde{a}_{m}^{e\sigma}\left(x_{3}\right)\hat{e}_{m\alpha}^{h\sigma}\right]\exp\left(ik_{m1}x_{1}+i\sigma k_{m3}x_{3}\right),\label{eq:H_pw}
\end{equation}
where the index $m$ enumerates a discrete set of plane waves having
wavevectors $\bm{k}_{m}^{\pm}=\left(k_{m1},0,\pm k_{m3}\right)$ defined
by the grating equation $k_{m1}=k_{1}^{inc}+2\pi m/\Lambda$, and
the dispersion equation $k_{m1}^{2}+k_{m3}^{2}=k_{b}^{2}$, $\Re k_{m3}+\Im k_{m3}\geq0$
with $k_{b}=\omega\sqrt{\varepsilon_{b}\mu_{0}}$. Unit vectors $\hat{\bm{e}}_{m}^{e\pm}$
and $\hat{\bm{e}}_{m}^{h\pm}$, which specify the polarization, explicitly
write
\begin{equation}
\begin{array}{c}
\hat{\bm{e}}_{m}^{e\pm}=\dfrac{\bm{k}_{m}^{\pm}\times\hat{\bm{e}}_{3}}{\left|\bm{k}_{m}^{\pm}\times\hat{\bm{e}}_{3}\right|},\\
\hat{\bm{e}}_{m}^{h\pm}=\dfrac{\left(\bm{k}_{m}^{\pm}\times\hat{\bm{e}}_{3}\right)\times\bm{k}_{m}^{\pm}}{\left|\left(\bm{k}_{m}^{\pm}\times\hat{\bm{e}}_{3}\right)\times\bm{k}_{m}^{\pm}\right|}.
\end{array}\label{eq:TEM-def}
\end{equation}
Then, the vertical coordinate dependent amplitudes occurring in Eqs. (\ref{eq:E_pw}),
(\ref{eq:H_pw}) become solutions of the following integral equations
(see \cite{Shcherbakov2018} for a detailed derivation)
\begin{equation}
\left(\!\! \begin{array}{c} \tilde{a}_{m}^{e+}\left(x_{3}\right) \\ \tilde{a}_{m}^{h+}\left(x_{3}\right) \end{array} \!\!\right) = \left(\!\! \begin{array}{c} \tilde{a}_{m}^{ext,e+}\left(x_{3}\right) \\ \tilde{a}_{m}^{ext,h+}\left(x_{3}\right) \end{array} \!\!\right) - \intop_{-\infty}^{x_{3}}d\zeta\dfrac{\mathcal{G}_{m}\left(x_{3},\zeta\right)}{2k_{m3}} \left(\!\! \begin{array}{c} \omega\mu_{b} \hat{\bm{e}}_m^{e+}\cdot{\bf J}_{m}\left(\zeta\right) - k_b \hat{\bm{e}}_{m}^{h+}\cdot{\bf M}_{m}\left(\zeta\right) \\ \omega\mu_{b} \hat{\bm{e}}_m^{h+}\cdot{\bf J}_{m}\left(\zeta\right) + k_b \hat{\bm{e}}_{m}^{e+}\cdot{\bf M}_{m}\left(\zeta\right) \end{array} \!\!\right)
\label{eq:sol_aep}
\end{equation}
\begin{equation}
\left(\!\! \begin{array}{c} \tilde{a}_{m}^{e-}\left(x_{3}\right) \\ \tilde{a}_{m}^{h-}\left(x_{3}\right) \end{array} \!\!\right) = \left(\!\! \begin{array}{c} \tilde{a}_{m}^{ext,e-}\left(x_{3}\right) \\ \tilde{a}_{m}^{ext,h-}\left(x_{3}\right) \end{array} \!\!\right) - \intop_{x_{3}}^{\infty} d\zeta\dfrac{\mathcal{G}_{m}\left(x_{3},\zeta\right)}{2k_{m3}} \left(\!\! \begin{array}{c} \omega\mu_{b} \hat{\bm{e}}_m^{e-}\cdot{\bf J}_{m}\left(\zeta\right) - k_b \hat{\bm{e}}_{m}^{h-}\cdot{\bf M}_{m}\left(\zeta\right) \\ \omega\mu_{b} \hat{\bm{e}}_m^{h-}\cdot{\bf J}_{m}\left(\zeta\right) + k_b \hat{\bm{e}}_{m}^{e-}\cdot{\bf M}_{m}\left(\zeta\right) \end{array} \!\!\right)
\label{eq:sol_ahm}
\end{equation}
with the coefficients $\mathcal{G}_{m}\left(\xi,\eta\right)$ coming from the plane wave decomposition of the Green's tensor. In general case of a planar waveguide they can be found, e.g., in \cite{Shcherbakov2017,Andreani2006},
and in particular, when $\varepsilon_{b}=\varepsilon_{1}=\varepsilon_{2}$
they explicitly read $\mathcal{G}_{m}\left(\xi,\eta\right)=\exp\left(ik_{3m}\left|\xi-\eta\right|\right)$ (note, that in this especial case the diffraction problem can still be nontrivial, once $\varepsilon(\bm{r})\neq\varepsilon_b$ in the grating region).
Here it is supposed that the sources are split in two parts: the first
part excites some known external fields $E_{\alpha}^{ext}$, $H_{\alpha}^{ext}$
with known plane wave decomposition amplitudes $\tilde{a}_{m}^{ext,e,h\pm}\left(x_{3}\right)$,
while the second part, $J_{\alpha}$, $M_{\alpha}$, present in the
integrands corresponds to some local sources to be specified below,
which bring an information about the mentioned difference between
the grating and the plane homogeneous layer in the region $D_{g}$.

\begin{figure}
\begin{centering}
\includegraphics[width=10cm]{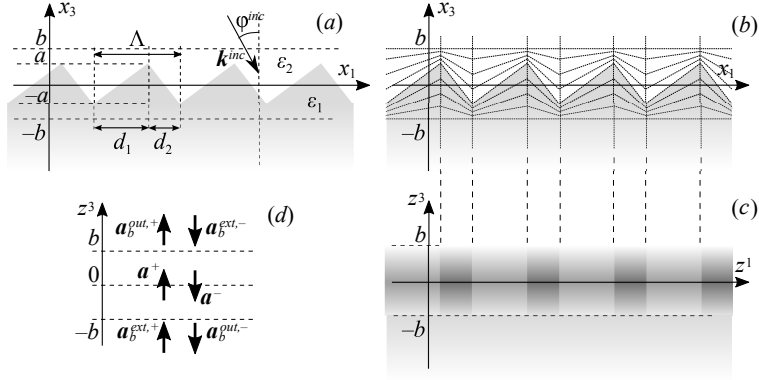}
\par\end{centering}
\caption{(a) Grating of period $\Lambda$, depth $2a$, and the grating region
$D_{g}=\left\{ \left|x_{3}\right|\protect\leq b:b\protect\geq a\right\} $;
(b) isolines of the curvilinear coordinates, which contain a plane
coinciding with the corrugation interface, and continuously become
the Cartesian at $x_{3}=\pm b$ boundaries; (c) illustration of an
effective periodic structure which can be thought to give rise to
the generalized metric sources; the structure has sharp vertical interfaces
when the grating corrugation profile has corners; (d) illustration to the divide-and-conquer approach used to efficiently calculate diffraction by metal gratings. The grating layer is split in two halves and self-consistent field amplitudes at $z^3=0$ are used to formulate the implicit Eq. (\ref{eq:smat_sc}).}
\label{fig1}
\end{figure}

\section{Generalized metric sources}
Once the general solution equations (\ref{eq:sol_aep})-(\ref{eq:sol_ahm})
are established they can be enclosed on the basis of the curvilinear
coordinate transformation idea mentioned in the introduction. Consider
curvilinear coordinates $\left(z^{1},z^{2},z^{3}\right)$ such that
in $D_{g}=\left\{ \left|x_{3}\right|\leq b:b\geq a\right\} $ the
coordinate plane $z^{3}=0$ coincides with the corrugation profile,
and the coordinates continuously become Cartesian at boundaries $\partial D=\left\{ x_{3}=\pm b\right\} $.
Such transformation can be defined as
\begin{equation}
\begin{array}{l}
x_{1,2}=z^{1,2},\\
x_{3}=\begin{cases}
z^{3}+\phi\left(z^{3}\right)f\left(z^{1}\right), & \left|x_{3}\right|\leq b\\
z^{3}, & \left|x_{3}\right|>b
\end{cases}
\end{array}\label{eq:transform_def}
\end{equation}
where $\phi\left(z^{3}\right)=1-\left|z^{3}\right|/b$. An
illustrative example is given in Fig. 1(b) . Possible generalizations
of the definition (\ref{eq:transform_def}) concern multilayer structures,
and are discussed in \cite{Shcherbakov2017} with the difference that
the corrugation function $f\left(x_{1}\right)$ is not required to
be smooth here. Transformation (\ref{eq:transform_def}) yields the
metric tensor
\begin{equation}
\left\{ g^{\alpha\beta}\right\} =\left(\!\!\!\begin{array}{ccc}
1 & 0 & -\dfrac{\phi f'}{1+\phi'f}\\
0 & 1 & 0\\
-\dfrac{\phi f'}{1+\phi'f} & 0 & \dfrac{1+\phi^{2}f'^{2}}{\left(1+\phi'f\right)^{2}}
\end{array}\!\!\!\right),\label{eq:metric_tensor}
\end{equation}
which components are discontinuous functions of coordinates due to the
discontinuity of $f'$.

Being written in the curvilinear coordinates the Maxwell's equations
become \cite{Schouten1989}
\begin{equation}
\begin{array}{c}
\xi^{\alpha\beta\gamma}\partial_{\beta}E_{\gamma}\left(\bm{r}\right)=i\omega\mu_{0}\sqrt{g}g^{\alpha\beta}H_{\beta}\left(\bm{r}\right)-M^{\alpha}\left(\bm{r}\right)\\
\xi^{\alpha\beta\gamma}\partial_{\beta}H_{\gamma}\left(\bm{r}\right)=-i\omega\varepsilon\left(\bm{r}\right)\sqrt{g}g^{\alpha\beta}E_{\beta}\left(\bm{r}\right)+J^{\alpha}\left(\bm{r}\right)
\end{array}\label{eq:Maxw_curvilinear}
\end{equation}
where the lower and upper indices distinguish the covariant and contravariant
vector components respectively, and $g$ denotes the determinant of
a matrix being inverse to (\ref{eq:metric_tensor}): $g=\det\left\{ g_{\alpha\beta}\right\} =1/\det\left\{ g^{\alpha\beta}\right\} $.
An observation that the metric tensor explicitly occurs only behind
the field terms in the right-hand parts of Eq. (\ref{eq:Maxw_curvilinear})
brings the second cornerstone idea of the GSMCC \cite{Shcherbakov2013}:
Eqs. (\ref{eq:Maxw_curvilinear}) can be rewritten in a form similar
to the Maxwell's equations in the Cartesian coordinates (\ref{eq:Maxw_cartesian})
and the rest can be attributed to generalized electromagnetic sources
originating from the difference between the curvilinear and the Cartesian
formulations. Namely, Eqs. (\ref{eq:Maxw_curvilinear}) become
\begin{equation}
\begin{array}{c}
\xi^{\alpha\beta\gamma}\partial_{\beta}E_{\gamma}\left(\bm{r}\right)=i\omega\mu_{b}\delta^{\alpha\beta}H_{\beta}\left(\bm{r}\right)-M_{gen}^{\alpha}\left(\bm{r}\right)-M^{\alpha}\left(\bm{r}\right),\\
\xi^{\alpha\beta\gamma}\partial_{\beta}H_{\gamma}\left(\bm{r}\right)=-i\omega\varepsilon_{b}\delta^{\alpha\beta}E_{\beta}\left(\bm{r}\right)+J_{gen}^{\alpha}\left(\bm{r}\right)+J^{\alpha}\left(\bm{r}\right)
\end{array}\label{eq:Maxw_curvilinear-gen}
\end{equation}
with the generalized metric sources
\begin{equation}
\begin{array}{c}
J_{gen}^{\alpha}\left(\bm{r}\right)=-i\omega\varepsilon_{b}\left(\eta_{e}\sqrt{g}g^{\alpha\beta}-\delta^{\alpha\beta}\right)E_{\beta}\left(\bm{r}\right),\\
M_{gen}^{\alpha}\left(\bm{r}\right)=-i\omega\mu_{b}\left(\eta_{h}\sqrt{g}g^{\alpha\beta}-\delta^{\alpha\beta}\right)H_{\beta}\left(\bm{r}\right).
\end{array}\label{eq:generalized_sources}
\end{equation}
Here $\eta_{e}=\varepsilon\left(z^3\right)/\varepsilon_{b}$, $\eta_{h}=\mu_{0}/\mu_{b}$.
Due to the similarity between the operator parts of Eqs. (\ref{eq:Maxw_cartesian})
and Eqs. (\ref{eq:Maxw_curvilinear-gen}) such decomposition allows
directly using solutions (\ref{eq:sol_aep})-(\ref{eq:sol_ahm}) with
the sources (\ref{eq:generalized_sources}) upon a mere substitution
of the coordinates $x_{\alpha}$ with $z^{\alpha}$ and recalling
that the modified fields should replace the real ones in Eq. (\ref{eq:generalized_sources}) according to Eq.~(\ref{eq:mod}).
The function $\varepsilon\left(\bm{r}\right)$ in the latter equations
depends only on the coordinate $z^{3}$ and takes two constant values
below and above the flattened corrugation interface: $\varepsilon\left(z^3\right)=\varepsilon_{1}$
for $z^{3}\leq0$, and $\varepsilon\left(\bm{r}\right)=\varepsilon_{2}$
for $z^{3}>0$, as Eq. (\ref{eq:transform_def}) states.

Eq. (\ref{eq:generalized_sources}) being substituted into (\ref{eq:sol_aep})-(\ref{eq:sol_ahm}) yields a self-consistent system, taking into account Eqs. (\ref{eq:mod})-(\ref{eq:H_pw}). At this point in case of a smooth coordinate transformation it is mathematically justified to perform the truncation of infinite Fourier series of the generalized sources \cite{Li1996,Chandezon1996}, so that Eqs. (\ref{eq:sol_aep})-(\ref{eq:sol_ahm}) become a finite linear integral equation system \cite{Shcherbakov2013}. In case of non-smooth transformations considered here a further effort is required, as the next section explains.

In analogy with the Generalized Source Method \cite{Shcherbakov2010,Shcherbakov2012}
and other volume integral methods in the Fourier space, which enclose
the volume integral equation with sources of the form ${\bf J}=-i\omega\left[\varepsilon\left(\bm{r}\right)-\varepsilon_{b}\right]{\bf E}$,
Eq. (\ref{eq:generalized_sources}) demonstrates that the curvilinear
metric impact can be treated as inhomogeneous and anisotropic material
tensors $\hat{\varepsilon}^{\alpha\beta}=\varepsilon\left(z^3\right)\sqrt{g}g^{\alpha\beta}$,
and $\hat{\mu}^{\alpha\beta}=\mu_{0}\sqrt{g}g^{\alpha\beta}$. Due
to the discontinuity of the metric tensor along the coordinate $z^{1}$
the points of discontinuity appear to effectively act like vertical
material interfaces, as Fig. 1(c) illustrates. Thus, special precautions
should be taken when calculating Fourier coefficients of the generalized
metric sources to be used in Eqs. (\ref{eq:sol_aep})-(\ref{eq:sol_ahm}),
in analogy with the Fourier methods in the Cartesian space \cite{Li1996}.
In case of the C-method the factorization rules were thoroughly explained
in \cite{Chandezon1996} in terms of relations between covariant field
components. The same rationale can be applied to the GSMCC, though,
here the correct Fourier space matrix-vector relations are demonstrated
to appear in a way analogous to the derivation of the electromagnetic
interface conditions, as the following section demonstrates.


\section{Fourier factorization of discontinuous metric sources}

The work \cite{Li1996} demonstrated that justifiability of truncation of infinite Fourier series of function products strongly depends on the continuity of the involved functions. Therefore, in case of discontinuous along $z^1$ metric tensor and field components, finite matrix relations between the Fourier components of the generalized sources and the fields should be properly derived. In order to demonstrate the correct truncated Fourier factorization consider the Maxwell's equations for the TE polarization (due to the presence of the both electric and magnetic sources, derivations for the TM polarization are quite similar):
\begin{equation}
\begin{array}{c}
\partial_{3}E_{2}=M^{1}-i\omega\mu_{b}H_{1},\\
\partial_{1}E_{2}=-M^{3}+i\omega\mu_{b}H_{3},\\
\partial_{3}H_{1}-\partial_{1}H_{3}=-i\omega\varepsilon_{b}E_{2}+J^{2},
\end{array}\label{eq:TE-eq}
\end{equation}
where the source terms represent a superposition of real sources exciting
incoming diffracting waves, and the local generalized metric sources.
In addition to Eqs. (\ref{eq:TE-eq}) let us explicitly recall the
Gauss law

\begin{equation}
\begin{array}{c}
\partial_{1}\left[\sqrt{g}\left(H_{1}+g^{13}H_{3}\right)\right]+\partial_{3}\left[\sqrt{g}\left(g^{31}H_{1}+g^{33}H_{3}\right)\right]=0,\\
\partial_{2}\left(\sqrt{g}E_{2}\right)=0.
\end{array}\label{eq:Gauss-TE}
\end{equation}
Here the dependency of $\varepsilon\left(\bm{r}\right)$ from the
coordinate $z^3$ only and $g^{11}=1$ property of Eq. (\ref{eq:metric_tensor})
were taken into account. The existence of the right-hand parts of Eqs.
(\ref{eq:TE-eq}), (\ref{eq:Gauss-TE}) requires the existence of
corresponding derivatives. Recalling the transition to the modified
fields, Eq.~(\ref{eq:mod}), and Eqs.~(\ref{eq:generalized_sources}), the last of Eq.~(\ref{eq:TE-eq}),
and the first of Eq. (\ref{eq:Gauss-TE}) become
\begin{equation}
\begin{array}{c}
\partial_{3}\tilde{H}_{1}-\partial_{1}C_{2}=-i\omega\varepsilon_{b}\tilde{E}_{2}+J^{2},\\
\partial_{1}C_{1}+\partial_{3}\tilde{H}_{3}=0,
\end{array}\label{eq:TE-modified}
\end{equation}
where the non-trivial combinations of the field and the metric tensor
components
\begin{equation}
\begin{array}{c}
C_{1}=\dfrac{1}{\sqrt{g}g^{33}}\tilde{H}_{1}+\dfrac{g^{13}}{g^{33}}\dfrac{1}{\eta_{h}}\tilde{H}_{3}\\
C_{2}=-\dfrac{g^{31}}{g^{33}}\tilde{H}_{1}+\dfrac{1}{\sqrt{g}g^{33}}\dfrac{1}{\eta_{h}}\tilde{H}_{3}
\end{array}\label{eq:C-def}
\end{equation}
should be continuous along $z^{1}$ coordinate, and, hence, across
the effective vertical interfaces discussed at the end of the previous
section. The continuity property follows directly from the same derivations
as the ones widely used in university textbooks to attain the electromagnetic
interface conditions by integrating Maxwell's equations in the vicinity of an interface, (see, e.g., \cite{Jackson1993}, Ch. 1.5.). Substitution
of the explicit $g^{\alpha\beta}$ components into Eq.~(\ref{eq:C-def})
yields
\begin{equation}
\begin{array}{c}
\left(\phi f'\right)C_{1}-C_{2}=-\left(1+\phi'f\right)\dfrac{1}{\eta_{h}}\tilde{H}_{3}\\
C_{1}+\left(\phi f'\right)C_{2}=\left(1+\phi'f\right)\tilde{H}_{1}
\end{array}\label{eq:C-g-explicit}
\end{equation}
Due to the continuity of $C_{1,2}$ and $(1+\phi'f)$ along $z^{1}$
one can directly take the Fourier transform of the latter equations \cite{Li1996} and pass to truncated series
to derive relations between the Fourier coefficients $C_{1,2m}$ and
$\tilde{H}_{1,3m}$. Denoting the Fourier component vectors with square
brackets as $\left[\bullet\right]$, Fourier-Toepltz matrices -- with
double square brackets as $\left\llbracket \bullet\right\rrbracket $;
expressing explicitly $\left[C_{1,2}\right]$ via $\left[\tilde{H}_{1,3}\right]$:
\begin{equation}
\left(\!\!\!\begin{array}{c} \left[C_1\right] \\ \left[C_2\right]\end{array}\!\!\!\right) = \left(\mathcal{I}+\mathcal{B}\mathcal{B}\right)^{-1}\left(\!\!\!\begin{array}{cc} \mathcal{I} & -\dfrac{1}{\eta_{h}}\mathcal{B}\\ \mathcal{B} & \dfrac{1}{\eta_{h}}\mathcal{I} \end{array}\!\!\!\right)\mathcal{A}\left(\!\!\!\begin{array}{c} \left[\tilde{H}_{1}\right]\\ \left[\tilde{H}_{3}\right] \end{array}\!\!\!\right),
\label{eq:C-H}
\end{equation}
with $\mathcal{A}=\left\llbracket 1+\phi'f\right\rrbracket $,
$\mathcal{B}=\left\llbracket \phi f'\right\rrbracket $, and identity matrix
$\mathcal{I}$; and substituting the modified fields into Eq. (\ref{eq:generalized_sources}),
one attains the following relations between the Fourier vectors of
the generalized metric sources and the fields:
\begin{equation}
\left(\!\!\!\begin{array}{c} \left[J^{2}\right] \\ \left[M^{1}\right] \\ \left[M^{3}\right] \end{array}\!\!\!\right) = -i\omega\epsilon_b \mathrm{M}^{-1}\mathrm{N} \left(\!\!\!\begin{array}{c} \left[\tilde{E}_{2}\right] \\ \left[\tilde{H}_{1}\right] \\ \left[\tilde{H}_{3}\right] \end{array}\!\!\!\right).
\label{eq:JM-EH}
\end{equation}
Here 
\begin{equation}
\mathrm{M} = \left(\!\!\!\begin{array}{ccc} \mathcal{I} & 0 & 0 \\ 0 & \mathcal{I}+\mathcal{B}\mathcal{B} & 0 \\ 0 & 0 & \mathcal{I}+\mathcal{B}\mathcal{B} \end{array}\!\!\!\right),
\label{eq:def-mat_M}
\end{equation}
\begin{equation}
\mathrm{N}^{e} = \left(\!\!\!\begin{array}{ccc}
\eta_{e}\mathcal{A}-\mathcal{I} & 0 & 0\\
0 & \eta_{h}\mathcal{A}-\mathcal{I}-\mathcal{BB} & -\mathcal{BA} \\
0 & -\mathcal{BA} & \mathcal{I}+\mathcal{BB}-\dfrac{1}{\eta_{h}}\mathcal{A} \end{array}\!\!\!\right).
\label{eq:def-mat_N}
\end{equation}
and diagonal matrix $\epsilon_b=diag\{\varepsilon_b,\,\mu_b,\,\mu_b\}$. The terms $\mathcal{BB}$, and $\mathcal{BA}$ imply that vectors should be successively multiplied by the corresponding matrices. Eqs.~(\ref{eq:C-H})-(\ref{eq:def-mat_N}) hold for each fixed value of $z^{3}$.



\section{Numerical method}
Definition of the coordinate transformation, Eq. (\ref{eq:transform_def}),
implies that the generalized metric sources (\ref{eq:generalized_sources})
are non-zero only when $\left|z^{3}\right|\leq b$. Therefore, the
integration limits in Eqs. (\ref{eq:sol_aep})-(\ref{eq:sol_ahm})
do not fall outside the region $D_{g}$. Let us introduce an equidistant
mesh (slicing) in $D_{g}$ defined by coordinates $z_{k}^{3}=\left(2k+1-N_{s}\right)b/N_{s}$,
$k=0,\dots,N_{s}-1$, with slice thickness $h_{s}=2b/N_{s}$, and
evaluate the integrals at the mesh points using the mid-point rule. Also denote  a maximum order of truncated infinite Fourier vectors and matrices as $N_{F}$.

Upon substitution of the generalized metric sources, Eqs.~(\ref{eq:sol_aep})-(\ref{eq:sol_ahm})
reduce to a set of linear equations on the unknown vector of
the TE wave amplitudes in each slice $\bm{a}_{mk}^{e}=\left(\tilde{a}_{m}^{e+}(z_{k}^{3}),\,\tilde{a}_{m}^{e-}(z_{k}^{3})\right)^{T}$:
\begin{equation}
\bm{a}^{e}=\bm{a}^{e,ext}+\mathrm{G}\mathrm{P}^{e}\mathrm{M}^{-1}\mathrm{N}^e\epsilon^{e}\bm{\bm{\mathcal{F}}}^{e}.
\label{eq:linear-equation}
\end{equation}
Here $\bm{a}^{e,ext}$ is the known amplitude vector of waves coming
from the exterior of the layer $D_{g}$, and the unknown field vector $\bm{\mathcal{F}}^{e}$ in case of the TE polarization is
\begin{equation}
\bm{\mathcal{F}}^{e} _{mk}=\left(\tilde{E}_{m2}(z_{k}^{3}),\,\tilde{H}_{m1}(z_{k}^{3}),\,\tilde{H}_{m3}(z_{k}^{3})\right)^{T}.
\label{eq:fvec}
\end{equation}
The Fourier index $m$ runs in range $-N_{F}\leq m\leq N_{F}$ and the spatial mesh index $k$ -- in range $0\leq k<N_{s}$. The block-diagonal matrix $\mathrm{P}^{e}$ defines the source-to-amplitude in-slice transformation of Eqs. (\ref{eq:sol_aep})-(\ref{eq:sol_ahm})
\begin{equation}
\mathrm{P}_{klmn}^{e}=\delta_{kl}\delta_{mn}\left(\!\!\!\begin{array}{ccc}
\dfrac{\omega\mu_{b}}{k_{m3}} & -1 & \dfrac{k_{m1}}{k_{m3}}\\
\dfrac{\omega\mu_{b}}{k_{m3}} & 1 & \dfrac{k_{m1}}{k_{m3}}
\end{array}\!\!\!\right),
\label{eq:def-mat-P}
\end{equation}
The matrix $\mathrm{G}$ is defined by operator $\mathcal{G}$ together with factor $-i\omega h_{s}/2$. In case $\varepsilon_{b}=\varepsilon_{1}=\varepsilon_{2}$
\begin{equation}
\mathrm{G}_{klmn}=-\delta_{mn}\frac{i\omega h_{s}}{2}\left(\!\!\!\begin{array}{cc}
\exp\left(ik_{m3}\left|z_{k}^{3}-z_{l}^{3}\right|\right) & 0\\
0 & \exp\left(ik_{m3}\left|z_{k}^{3}-z_{l}^{3}\right|\right)
\end{array}\!\!\!\right).
\label{eq:def-mat-G}
\end{equation}
Generally, when the two permittivities $\varepsilon_{1,2}$ present in
the regions $x_{3}<-b$ and $x_{3}>b$ are different from $\varepsilon_{b}$,
the Green's tensor becomes more complex, and the latter matrix
element should be replaced with another one incorporating multiple
reflections at interfaces $x_{3}=\pm b$. For explicit equations in
this case see, for example, \cite{Shcherbakov2017,Andreani2006}.

Multiplication of Eq.~(\ref{eq:linear-equation}) by
\begin{equation}
\mathrm{Q}_{klmn}^{e}=\delta_{kl}\delta_{mn}\left(\begin{array}{cc}
-1 & -1 \\ \dfrac{k_{m3}}{\omega\mu_{b}} & -\dfrac{k_{m3}}{\omega\mu_{b}}\\
-\dfrac{k_{m1}}{\omega\mu_{b}} & -\dfrac{k_{m1}}{\omega\mu_{b}}
\end{array}\right),
\label{eq:def-mat-Q}
\end{equation}
which is composed of components of unit vectors (\ref{eq:TEM-def}), yields the desired linear equation system
\begin{equation}
\bm{\mathcal{F}}^{e} = \bm{\mathcal{F}}^{e,ext}+\mathrm{Q}^e\mathrm{G}\mathrm{P}^{e}\mathrm{M}^{-1}\mathrm{N}^e\epsilon^{e}\bm{\mathcal{F}}^{e}.
\label{eq:linear-equation-F}
\end{equation}
The derivation of Eq.~(\ref{eq:linear-equation-F}) is analogous
to what is done in \cite{Shcherbakov2013,Shcherbakov2017}, but here
one faces with an additional matrix inversion coming from the field-to-source transformation of Eq.~(\ref{eq:JM-EH}). Hereof it follows, that when solving this system by an iterative method, each iteration would require performing a numerically expensive operation of matrix inversion. This, in turn, would result in total $O(N_sN_F^3)$ numerical complexity instead of the GSMCC breakthrough $O(N_sN_F\log(N_sN_F))$ \cite{Shcherbakov2013}, which is based on a decomposition of the corresponding linear system matrix into block-diagonal and block-Toeplitz constituents, and application of the Fast Fourier Transform for speeding-up multiplications. To handle this inversion let us multiply Eq.~(\ref{eq:linear-equation-F}) by $\mathrm{M}^{-1}\mathrm{N}^e\epsilon^e$ and introduce the new vector
\begin{equation}
\bm{\bm{\mathcal{J}}}^{e}=\mathrm{M}^{-1}\mathrm{N}^e\epsilon^{e}\bm{\bm{\mathcal{F}}}^{e}=\left(\mathrm{M} - \mathrm{N}^e \epsilon^{e} \mathrm{Q}^{e}\mathrm{G}\mathrm{P}^{e}\right)^{-1}\mathrm{N}^e\epsilon^{e}\bm{\bm{\mathcal{F}}}^{e,ext}.\label{eq:linear-system}
\end{equation}
This system is solved for the unknown vector $\bm{\bm{\mathcal{J}}}^{e}$,
it can be substituted into Eq.~(\ref{eq:linear-equation-F})
to attain a vector of diffracted amplitudes at the grating region boundaries:
\begin{equation}
\bm{a}^{e,out}_b=\bm{a}_{b}^{e,ext}+\mathrm{G}_{b}\mathrm{P}^{e}\bm{\bm{\mathcal{J}}}^{e},
\label{eq:solution-external}
\end{equation}
Here the output and external amplitude vectors are $\bm{a}_{b,m}^{e,out/ext} = \left(\tilde{a}_{m}^{out/ext,e+}(b),\,\tilde{a}_{m}^{out/ext,e-}(-b)\right)^{T}$,
and the matrix operator $\mathrm{G}_{b}$ also comes from tensor $\mathcal{G}$,
and ``collects'' the waves diffracted in each slice at $D_{g}$
boundaries. In case $\varepsilon_{b}=\varepsilon_{1}=\varepsilon_{2}$
\begin{equation}
\mathrm{G}_{b,kmn}=-\delta_{mn}\frac{i\omega h_{s}}{2}\left(\!\!\!\begin{array}{cc} \exp\left(ik_{m3}\left|b-z_{k}^{3}\right|\right) & 0 \\ 0 & \exp\left(ik_{m3}\left|z_{k}^{3}+b\right|\right) \end{array}\!\!\!\right).
\end{equation}
Since the curvilinear metric continuously transforms to
the Cartesian one at $\left|z^{3}\right|=b$, and the modified field
coincides with the real one outside the grating region, the output
amplitudes immediately define the diffracted field outside $D_{g}$,
and no other transformation is required. For the TM polarization resulting equations are the same as Eqs. (\ref{eq:linear-system}), (\ref{eq:solution-external}), but with slightly different matrices $\mathrm{N}$, $\epsilon$, $\mathrm{P}$, and $\mathrm{Q}$. They are listed in Appendix A.

The provided formulation of Eqs. (\ref{eq:linear-system}), (\ref{eq:solution-external}) theoretically preserves the fast and memory efficient conception of the previous papers \cite{Shcherbakov2013,Shcherbakov2017}, and allows one to
perform the grating diffraction calculation in $O\left(N\log N\right)$
time and $O\left(N\right)$ memory resort with $N=N_{s}N_{F}$. Namely, this comes from the fact that multiplications by block-diagonal matrices $\mathrm{P}^{e}$, $\mathrm{Q}^{e}$,
$\mathrm{G}$, and $\mathrm{G}_{b}$ are of linear asymptotic complexity,
and multiplications by $\mathrm{M}$, and $\mathrm{N}^e$ are of $O\left(N\log N\right)$ asymptotic complexity due to the block-Toeplitz
structure.

In order to demonstrate the validity of the developed method and codes as well as to study the asymptotic rate of convergence in dependence of $N_{F}$ let us consider an example of a non-symmetric triangular corrugation profile with the period-to-wavelength ratio $\Lambda/\lambda=1.5$ and $d_1/d_2 = 2/3$ (see Fig.~\ref{fig1}(a)). Matrices (\ref{eq:def-mat_M}), (\ref{eq:def-mat_N}) are found analytically in case of an arbitrary piecewise-linear finction $f$. For convergence calculation there was chosen the starting value $N_{F,0} = 16$, and a set of method runs for subsequent values of $N_{F}$ was performed. With the increase of $N_{F}$ the size of output amplitude vectors $\bm{a}^{out}$ also increases. Therefore, only central amplitudes with indices $-N_{F,0}\leq m\leq N_{F,0}$ were picked up for comparison. The convergence was plotted as the dependence of the maximum absolute value $\max|a^{out}_m(N_{F,k+1})-a^{out}_m(N_{F,k})|$, $N_{F,k+1}=N_{F,k}+1$, from the inverse of $N_{F,k}$. Since the validity of the former GSMCC formulation was established previously \cite{Shcherbakov2013}, and generally incorrect and correct formulations of the Fourier methods are known to converge to the same results \cite{Granet1996,Li1996}, no other reference method for calculation of diffraction amplitudes is used here. Other polygonal grating shapes can be simulated by means of a sample Matlab code downloadable from \cite{gsmcc}.

\begin{figure}
\begin{centering}
\includegraphics[width=9cm]{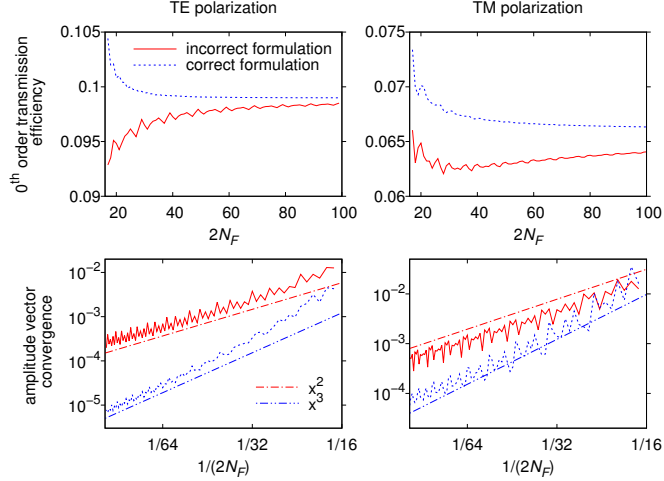}
\par\end{centering}
\caption{Convergence of the GSMCC in the previous formulation, which is incorrect for gratings having corrugation profile corners (red solid line),
and in the present correct formulation (blue dashed line). The graphs
are calculated for a triangular dielectric grating with $d_1/d_2 = 2/3$ (see Fig.~\ref{fig1}), depth-to-wavelength ratio $2a/\lambda=1.0$, the period-to-wavelength ratio $\Lambda/\lambda=1.5$, and the angle of incidence $10^{\circ}$. The number of slices is $N_{s}=1000$, permittivities of the covering medium and the grating are $\varepsilon_{1}=1$, $\varepsilon_{2}=\left(2.5\right)^{2}$ respectively. The lower pair of graphs demonstrates the convergence of the central part of diffraction amplitude vector $\bm{a}^{e,out}$ corresponding to the lowest used number of Fourier harmonics $2N_{F,0} = 16$.}
\label{fig:conv_d}
\end{figure}

\begin{figure}
\begin{centering}
\includegraphics[width=9cm]{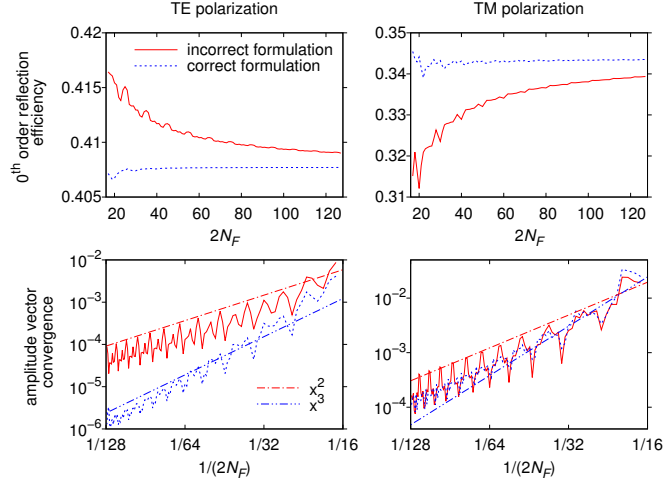}
\par\end{centering}
\caption{Same as in Fig. 2 but for metallic grating of permittivity $\varepsilon_{2}=\left(0.2+3.2i\right)^{2}$ and $2a/\lambda=0.5$.}
\label{fig:conv_m}
\end{figure}

Examples of the method convergence in dependence of $N_{F}$ are given
in Figs.~\ref{fig:conv_d},\ref{fig:conv_m} for the mentioned triangular corrugation profile analysed
with the formulation of \cite{Shcherbakov2013}, which is incorrect
for grating with corners, and with the correct formulation provided
here. Grating depth-to-wavelength ratio, and slice thickness were taken to be $2a/\lambda=1.0$, $h_s = 0.001\lambda$ for a dielectric grating with $\varepsilon_{1}=1$, $\varepsilon_{2}=\left(2.5\right)^{2}$, and $2a/\lambda=0.5$, $h_s = 0.0005\lambda$ for a metallic grating with the same $\varepsilon_{1}$ and $\varepsilon_{2}=\left(0.2+3.2i\right)^{2}$. The angle of incidence was $10^{\circ}$. The upper subplots in each figure show changes in absolute values of zero order diffraction efficiencies, and the lower subplots demonstrate the convergence defined above. The trendlines $x^2$ and $x^3$, reveal, that the new formulation not only allows obtaining substantially more accurate results for a given $N_F$ than the old one, but also that it exhibits a superior rate of convergence -- close to cubic but in the case of the metallic grating for the TM polarization, which might be related to a poor Fourier space field representation in the vicinity of metallic corners. It can be also noticed that the convergence of the incorrect formulation in case of the TM polarization is rather slow.

\begin{figure}
\begin{centering}
\includegraphics[width=10cm]{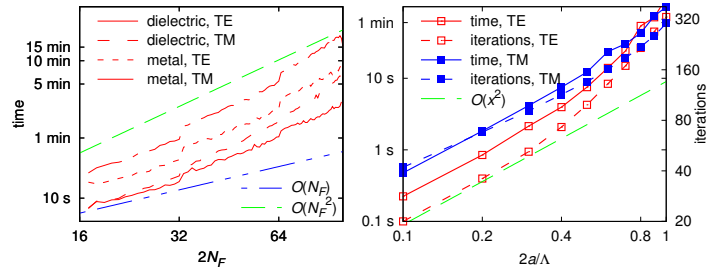}
\par\end{centering}
\caption{Left figure: dependence of the GSMCC run time from the maximum absolute value of the Fourier harmonic number $N_F$ for the four considered  diffraction problems corresponding to Figs. \ref{fig:conv_d},\ref{fig:conv_m}. Right figure: Dependence of the calculation time and the number of GMRes iterations from the depth-to-period ratio in case of dielectric gratings with all parameters, but depth, being the same as in Fig. \ref{fig:conv_d}.}
\label{fig:time}
\end{figure}

Despite the theoretical complexity of the method is close to linear, in practice it appears to get worse due to an impact of a linear equation iterative solver. The presented examples were calculated by means of the Generalized Minimal Residual method (GMRes). Fig.~\ref{fig:time} shows the dependence of the method run time from $N_F$ for the four considered cases of the dielectric and metal gratings under the TE and TM polarizations. The run time appears to grow linearly for small $N_F$, and faster for larger numbers of the Fourier harmonics, which is due to the increase of a number of iterations required for the GMRes to converge down to a given tolerance (taken to be $10^{-7}$ here). In case of the dielectric grating the iteration number grew from about 100 to 350. In case of metallic gratings it was found that solution of Eq.~(\ref{eq:linear-system}) for the whole grating exhibits quite slow GMRes convergence, or even stagnation. To solve this issue, the divide-and-conquer strategy described in Section~6 of \cite{Shcherbakov2017} was applied (see also \cite{Iff2017}). A particular case of this strategy implemented in this work is outlined in Appendix B.

For practical applications it is also important to track a dependence of the run time from grating depth-to-period ratio. Such dependence is shown in the right-hand part of Fig.~\ref{fig:time} for dielectric gratings of the same geometry as considered above. For this set of simulations with increasing depth the vertical spatial resolution remained constant, $h_s = 10^{-3}\lambda$, hence the number of slices $N_s$ increased linearly with the ratio $2a/\Lambda$. The dependence appears to be nonlinear as the number of GMRes iterations, shown in the right vertical axis, also increased. A corresponding time-depth dependence in case of metal gratings is quite similar.

\section{Conclusion}
To summarize this work extends the applicability of the GSMCC method by providing an explicit formulation of the method in case of gratings having corrugations profiles with corners. Eqs. (\ref{eq:linear-system}), (\ref{eq:solution-external}) reveal that the previously demonstrated low asymptotic numerical complexity and memory consumption can be preserved in the new formulation, and the numerical examples show its importance in attaining accurate diffraction efficiencies. The presented theory can be also applied in the case of conical diffraction by simply extending the involved matrices \cite{Shcherbakov2010}. Calculation of diffraction by crossed gratings would require a separate revision of matrix-vector products, and will be reported elsewhere.

Concerning "overhanging" gratings, which profiles cannot be defined by single-valued functions, a parametric transformation can be used leading to a metric tensor being very similar to Eq. (\ref{eq:metric_tensor}). The rest formulation would remain the same. An evidence for the GSMCC to be capable to handle such cases was reported in \cite{Maurel2015}, where a curvilinear Fourier modal method was demonstrated to work for "overhanging" structures. This means that, in particular, the method should be valid for vertical wall structures. The main issue is either to find an appropriate transformation allowing to analytically calculate necessary Fourier matrices, or to develop a suitable method for numerical calculation of these matrices. In any case such demonstration lies outside the scope of this paper, and might be a subject of a future publication.

Despite the demonstrated practical calculation time growth with the increase of $N_F$ is faster than linear, it is still less than the cubic dependence inherent to the modal approaches. Absolute values of run times are strongly dependent on a programming implementation and hardware, and can be substantially decreased by using the graphical processing units. Therefore, the attained results can be further utilized in the field of grating structure optimization, where computationally efficient and reliable rigorous solvers of the Maxwells equations are of great importance.

\section*{Funding}
Russian Science Foundation (RSF) (17-79-20345).


\section*{Appendix A}

In case of the TM polarization solution of the diffraction problem is still given by Eqs. (\ref{eq:linear-system}), (\ref{eq:solution-external}), but with slightly different matrices $\mathrm{N}^h$, $\epsilon^h$, $\mathrm{P}^h$, and $\mathrm{Q}^h$. Namely, $\mathrm{N}^h$ is the same as Eq.~(\ref{eq:def-mat_N}), but with $\eta_e$ and $\eta_h$ interchanged. The other matrices are
\begin{equation}
\epsilon_{mn,kl}^{h}=\delta_{mn}\delta_{kl} diag\left\{ \mu_{b},\varepsilon_{b},\varepsilon_{b}\right\},
\label{eq:matepsh}
\end{equation}
\begin{equation}
\mathrm{P}_{klmn}^{h}=\delta_{kl}\delta_{mn}\left(\!\!\!\begin{array}{ccc}
\dfrac{\omega\varepsilon_{b}}{k_{m3}} & 1 & -\dfrac{k_{m1}}{k_{m3}}\\
\dfrac{\omega\varepsilon_{b}}{k_{m3}} & -1 & -\dfrac{k_{m1}}{k_{m3}}
\end{array}\!\!\!\right),
\label{eq:def-mat-P-1}
\end{equation}
and
\begin{equation}
\mathrm{Q}_{klmn}^{h}=\delta_{kl}\delta_{mn}\left(\begin{array}{cc}
-1 & -1\\
-\dfrac{k_{m3}}{\omega\varepsilon_{b}} & \dfrac{k_{m3}}{\omega\varepsilon_{b}}\\
\dfrac{k_{m1}}{\omega\varepsilon_{b}} & \dfrac{k_{m1}}{\omega\varepsilon_{b}}
\end{array}\right).
\label{eq:def-mat-Q-1}
\end{equation}

\section*{Appendix B}

As noted in Section 5, application of Eq.~(\ref{eq:linear-system}) to a whole metallic grating results in a poor convergence or even stagnation of the GMRes depending of grating depth. A remedy to this issue is the use of a divide-and-conquer strategy described in Section~6 of \cite{Shcherbakov2017}. To briefly outline it, denote the diffraction operator explicitly given by Eqs. (\ref{eq:linear-equation-F}), (\ref{eq:solution-external}) as $\mathcal{S}$:
\begin{equation}
\bm{a}^{out}_b = \mathcal{S}(-b,b)*\bm{a}^{ext}_b \Leftrightarrow \left(\!\!\! \begin{array}{c} \bm{a}^{out-}_b \\ \bm{a}^{out+}_b \end{array} \!\!\!\right) = \left(\!\!\! \begin{array}{cc} \mathcal{S}^{-+}(-b,b) & \mathcal{S}^{--}(-b,b) \\ \mathcal{S}^{++}(-b,b) & \mathcal{S}^{+-}(-b,b) \end{array} \!\!\!\right) * \left(\!\!\! \begin{array}{c} \bm{a}^{ext+}_b \\ \bm{a}^{ext-}_b \end{array} \!\!\!\right)
\label{eq:smat}
\end{equation}
The dependence on the lower and the upper bounds of the grating layer $\pm b$ emphasizes the fact that the operator is applied to the whole grating layer. When the layer is too thick for the linear iterative solver to converge in reasonable time, this layer can be divided in two as Fig.~\ref{fig1}d illustrates. Let us introduce intermediate self-consistent amplitude vectors between the layers $\bm{a}^{\pm}(z^3=0)$. Then, defining the diffraction operators of the lower and the upper half-layers as $\mathcal{S}(-b,0)$, and $\mathcal{S}(0,b)$ allows to write out equations similar to Eq. (\ref{eq:smat}) for the halves including the intermediate amplitude vectors. Rearrangement of blocks of these equations yields
\begin{equation}
\left(\!\!\! \begin{array}{cc} \mathcal{I} & -\mathcal{S}^{+-}(-b,0) \\ -\mathcal{S}^{-+}(0,b) & \mathcal{I} \end{array} \!\!\!\right) * \left(\!\!\! \begin{array}{c} \bm{a}^{+}(z^3=0) \\ \bm{a}^{-}(z^3=0) \end{array} \!\!\!\right) = \left(\!\!\! \begin{array}{c} \mathcal{S}^{++}(-b,0) * \bm{a}^{ext+}_b \\ \mathcal{S}^{--}(0,b) * \bm{a}^{ext-}_b \end{array} \!\!\!\right)
\label{eq:smat_sc}
\end{equation}
This equation system can be in turn solved by the GMRes. Therefore, solutions of the linear equation systems become nested: while solving Eq. (\ref{eq:smat_sc}), each iteration requires two solutions of Eqs. (\ref{eq:linear-equation-F}), (\ref{eq:solution-external}) for each half-layer. Once vectors $\bm{a}^{\pm}(z^3=0)$ are calculated, the required output amplitudes come from
\begin{equation}
\begin{split}
\bm{a}_b^{out+} &= \mathcal{S}^{++}(0,b) * \bm{a}^{+}(z^3=0) + \mathcal{S}^{+-}(0,b) * \bm{a}^{ext,-}_b ,\\
\bm{a}_b^{out-} &= \mathcal{S}^{--}(-b,0) * \bm{a}^{-}(z^3=0) + \mathcal{S}^{-+}(-b,0) * \bm{a}^{ext,+}_b.
\end{split}
\label{eq:smat_out}
\end{equation}
This approach can be generalized to any number of sub-layers of the initial grating layer. Such implicit self-consistent method allows to dramatically decrease calculation time for metal gratings providing that all plane material interfaces in the curvilinear coordinates coincide with some sub-layer boundaries, and constant $\varepsilon_b$ is chosen to be different for each sub-layer, being equal to an averaged sub-layer permittivity.

\bibliography{gsmccd}
\bibliographystyle{ieeetr}
\end{document}